\documentclass{eas}
\usepackage{graphicx}
\newcommand{\HII}{H\,{\sc ii}}

\newcommand{\kms}{km~s$^{-1}$}

\newcommand{\Ls}{$L_{\odot}$}
\newcommand{\vinfty}{$v_{\infty}$}
\newcommand{\Md}{$\dot{M}$}
\newcommand{\nv}{N\,{\sc v} $\lambda$1238-42}
\newcommand{\siiv}{Si\,{\sc iv} $\lambda$1393-1402}
\newcommand{\civ}{C\,{\sc iv} $\lambda$1548-51}

\newcommand{\siv}{S\,{\sc iv} $\lambda$1062-72}
\newcommand{\svi}{S\,{\sc vi} $\lambda$933-44}
\newcommand{\pv}{P\,{\sc v} $\lambda$1118-28}
\newcommand{\ciii}{C\,{\sc iii} $\lambda$977}
\newcommand{\niii}{N\,{\sc iii} $\lambda$989-91}
\newcommand{\Teff}{$T_{\rm{eff}}$}

\begin{document}

%%-----------------------------
%%      the top matter
%%-----------------------------
\title{Fundamental Parameters of Massive Stars} 
\author{Paul A. Crowther}\address{Dept. of Physics \& Astronomy\\
University College London\\Gower Street\\London WC1E 6BT}
%\author{...}\address{...}
%\author{...}\address{...}
%
\runningtitle{Crowther: Fundamental Parameters of Massive Stars}
\begin{abstract}
We discuss the determination of fundamental parameters of 
`normal' hot, massive OB-type stars, namely
temperatures, luminosities, masses, gravities and surface abundances. 
We also  present methods used to derive properties of stellar winds 
-- mass-loss rates and wind velocities from early-type stars.
\end{abstract}
\maketitle
%%-----------------------------
%%      your text
%%-----------------------------
\section{Introduction}

The vast majority of stars in the Milky Way and external galaxies
have masses close to that of the Sun, and live long ($\geq$Gyr),
relatively peaceful lives. In contrast, stars with initial masses
in excess of $\approx$10$M_{\odot}$, with
O-type and early B-type spectral types on the main-sequence
 live for a relatively short life
($\sim$5--20 Myr), though 
have an astrophysical importance which belies their
rarity. Massive stars and their evolved descendants possess powerful stellar
winds, such that they are important contributors to the chemical 
and dynamical evolution of galaxies, due to their influence on the 
composition and energetics of 
the interstellar medium (ISM). Indeed, the primary origin of oxygen
and other $\alpha$-elements in the universe is from massive stars.
Ultimately, they end their 
life as a core-collapse Supernovae, causing further `feedback' to the ISM.

Massive stars spend the majority of their lives in the blue part of the
Hertzsprung-Russell (HR) diagram, and so emit the majority of their energy
in the far- and extreme- ultraviolet. Consequently, in normal galaxies they
are the dominant source of Lyman continuum photons. A typical O7 dwarf emits
10$^{49}$ Lyman continuum photons every second for its main sequence life,
and so OB stars ionize material in their vicinity, producing ionized nebulae,
or \HII regions. Within the Milky Way, the Orion Nebula is the 
closest birth place of massive stars. Further afield, 30 Doradus, the
Tarantula Nebula represents the greatest single concentration of massive
stars within the Local Group of galaxies, containing many hundreds of 
OB stars within its Giant \HII region. 
Further afield, starburst galaxies contain many so-called `super-star
clusters'.  Massive stars are the only stellar objects
to be {\it observed} in high redshift galaxies (e.g. MS 1512-cB58:  Pettini
et al. 2000)

It is clear that massive stars play a major role in astrophysics. In this
article I shall explain how one can determine the fundamental properties of 
main sequence massive stars. In the following article, I shall discuss 
post-main sequence massive stars and their end states.

%Nevertheless, a caveat exists on their
%astrophysical and cosmological applications: in order to properly employ 
%massive stars as useful probes, one need to accurately determine and 
%model their spectral energy distribution and their mass-loss properties
%as a function of metallicity.

\section{Basic Concepts}

The principal measured parameter of an individual star is its 
brightness in the night sky, and its colour, namely the difference
between magnitudes at two or more wavelengths. Historically, 
photography was the method of choice for astronomers. Since photography
is most sensitive to visible radiation, spectral classification 
traditionally concentrated on these wavelengths. 
Only recently have electronic detectors, such as CCDs supplanted 
photographic plates.

Brightness, colour, and spectral types can lead to the luminosity and
temperature of the star, the fundamental parameters to place it on the
Hertzsprung-Russell diagram.
The apparent magnitude of a star is typically measured 
through a filter, commonly the $V$ defined in the Johnson $UBV$ system. 
The absolute visual magnitude, $M_{V}$ is related to the apparent
magnitude, $m_{V}$ via the relationship, 
\[ M_{V} = m_{V} - 5 + 5 \log d - A_{V} \]
where $d$ is the distance (in parsec) and $A_{V}$ is the $V$-band extinction,
defined by $A_{V} = R \times E(B-V)$, where $R\sim$3.1. $E(B-V)$ is merely
the difference between the {\it observed} colour excess $m_{B} - m_{V} = B-V$ 
in the Johnson system and the {\it intrinsic} $(B-V)_0$ colour. 
According to Wien's law, most of the stellar
radiation for luminous hot stars with their $\approx 30,000~$K minimum
\Teff~ in emitted in the far UV regions.
At optical wavelengths one is
observing way out on the long wavelength tail of the spectral energy
distribution (SED) for hot stars. In
the $UBV$ system, intrinsic colors of OB stars are
thus nearly degenerate (e.g., differing only by a few hundredths of a
magnitude), typically $(B-V)_0 \approx -$0.3 mag.

The highly wavelength dependent extinction of starlight by interstellar
dust dominates the colors of luminous stars in the Milky Way. The observed
optical colors of hot stars are used primarily to determine properties of
the intervening interstellar medium and are {\it not} very useful to
evaluate stellar parameters.
Using the standard Galactic extinction curve of Seaton (1979), 
for every magnitude of extinction
at the $V$-band, the far-UV extinction at 1200\AA\ suffers four magnitudes,   
whilst only 0.4 magnitudes is suffered at 1$\mu$m and 0.1 magnitudes at
10$\mu$m. Consequently, regions, such as the Galactic Centre which are
invisible at UV and optical wavelengths are reasonably transparent longward
of the $K$-band. Internal extinction in other galaxies differs somewhat
from the Milky Way (Howarth 1983; Bouchet et al. 1985). In addition to 
continuous dust absorption, there are also discrete interstellar absorption
features, principally  the atomic Lyman HI and molecular H$_2$ bands at UV
wavelengths, whose strengths broadly scale with $E(B-V)$ (see  e.g.
Pellerin et al. 2002)

The $m_V$ measures the brightness of a star through a filter, 
but a luminosity is the emergent radiation over all wavelength bands.
The Sun has an absolute visual magnitude of $M_{V}$=4.82 mag, whilst
typical luminous hot stars have $M_{V} = - 5$ mag, so they are intrinsically
of order 10,000 times brighter in the visual. Considering their output over
all wavelengths, they exceed this by a further factor of 10 to 100. In magnitudes, one can write, $M_{\rm bol} = M_{V} + B.C$, where $M_{\rm bol}$ is
the {\it bolometric} magnitude and $B.C.$ is the bolometric {\it correction}.
Thus latter quantity is strongly dependent on the $T_{\rm eff}$, having
an appreciable value for hot, blue stars. One can relate the luminosity of
a star in solar luminosities to $M_{\rm bol}$ via
\[ M_{\rm bol} = -2.5 \log (L/L_{\odot}) + 4.74 \]
where $L_{\odot} = 3.845 \times 10^{33}$ erg\,s$^{-1}$.

In principal, one obtains the luminosity of a star by measuring
its $m_{V}$ and $(B-V)$, obtaining a distance, adopting a $B.C.$ and
utilizing the above equations together with a calibration relating the
spectral type of the star to its intrinsic $B-V)_0$ colour. A $T_{\rm eff}$
will first need to be obtained so that the $B.C.$ can be determined.
Various $B.C.(T_{\rm eff})$ exist in the literature (e.g. Balona 1994). 
For O and early B stars, Vacca et al. (1996) derived 
\[ B.C.  = 27.66 - 6.84 \log T_{\rm eff} \]
The dependence is clearly very steep for O stars, so small errors in $T_{\rm eff}$ may lead to large errors in luminosity. 
For example, the stellar luminosity
of an O3 dwarf with absolute magnitude $M_{\rm V} = -5$ mag is 500,000$L_{\odot}$ if \Teff=50,000K, but only 370,000\Ls, if \Teff=45,000K.

It is, of course, no trivial matter to
determine the distance of an individual hot star, and this represents
the greatest difficulty with deriving luminosities for Galactic OB stars.
 Due to their large distances, direct determination via stellar parallax
methods are only possible for a dozen cases with Hipparcos (Lamers
et al. 1997). Amongst the closest examples Hipparcos measured were 
$\zeta$ Oph (O9.5~Vn) and $\tau$ Sco (B0.2~IV) 
at distances of $\sim$135 pc, a hundred times more distant from the 
Sun than the closest star, $\alpha$ Centauri. 

Within a decade, {\sc gaia} should enable reliable parallaxes to all
optically visible Galactic OB stars. Until then, one has to
resort to obtaining 
spectral type-$M_{\rm V}$ calibrations using OB stars in associations or
clusters. Humphreys (1978) carried out a very extensive study for Galactic
OB supergiants. This was 
extended to Magellanic Cloud associations by P. Massey and co-workers. 
These calibrations may then be applied to individual stars in the field.
The advantage of Magellanic Cloud stars is that their distances  are
known to a precision better than 10\%. Consequently, 
despite their relative proximity,  luminosities of Galactic OB stars 
are more imprecise
than those of stars in the Magellanic Clouds or beyond. The limitation
of extragalactic stars is principally that of spatial resolution.
Fortunately, one finds reasonably good consistency between 
calibrations from Galactic and extra-galactic hot stars. Nevertheless,
OB stars within a given spectral type show a substantial spread, which
is thought to be typically
$\pm$0.5 mag. 
%Hipparcos showed that differences may 
%exceed one magnitude, corresponding to a factor of $>$2.5 in 
%luminosity, with slowly rotating OB stars 
%systematically fainter in $M_{\rm V}$ than
%rapid rotators of the same spectral type.

%HR diagram. magnitude scale, bolometric correction, distances (Hipparcos).
%Extinction.

One may also use radio observations to study the Galactic distribution of
massive stars, since these are located in Giant \HII regions that are
strong radio emitters (e.g. Smith et al. 1978). OB stars are confined
to the thin disk of the Milky Way, and (crudely) map out the spiral arm
structure (Georgelin \& Georgelin 1976). 
Knowledge of the ionizing fluxes of individual OB stars 
permits one to obtain total numbers of stars in \HII regions. The ionizing
fluxes of hot stars is a very strong function of temperature and to
a lesser degree surface gravity. Vacca et al. (1996) provided a $T{\rm eff}$
calibration of Lyman ($Q_{0}$) and He\,{\sc i}($Q_{1}$) 
ionizing fluxes for OB stars. 
The widely
used $T_{\rm eff}$--spectral type relation for OB stars adopted
therein (see below) has been substantially revised recently, such
that  Crowther \& Dessart (1998) provide  a calibration of 
$Q_{0}(T_{\rm eff}, \log L)$, 
whilst Hubeny \& Lanz (2003) have recently produced a comprehensive tabulation
of ionizing fluxes for dwarf O stars at a range of metallicities and gravities.

\section{Spectral Types}

The emergent continuum radiation from stars arises from the stellar
photosphere while the absorption line spectrum comes from the atmosphere.  
In normal stars these regions overlap, can be treated as plane parallel,
and are closely coupled to the stellar parameters.

Numerical subtypes
distinguish differences within the spectral letters and go from number 0
to 9 (but with a few 0.5 divisions). Line widths
and other line ratios are used to 
determine the atmospheric pressures (related to
gas densities, local gravity, hence luminosity). These labels utilize
roman numerals I--V, with the latter corresponding to the faintest
luminosity class. While five luminosity classes are easily distinguished
among cool K and M stars with their wide dispersion in $M_V$, only I, III,
and V are adopted for O type stars which have a more limited $M_V$ range.

The most common system of spectral classification builds upon that of
Morgan et al. (1943--MK) in which moderate spectral resolution 
is used to define natural groups of stars with similar spectral
characteristics.  This two dimensional system uses selected letters and
numbers for the spectral types and subtypes, and roman numerals for the
luminosity classes. {\it Standard stars} are typically selected for each
spectral type and luminosity class. Classification of other stars is then
carried out by obtaining their spectra and by using the spectral criteria
an eye comparison to the standards. The dividing line for luminous hot
stars at \Teff~ just below $30000~$K corresponds roughly to spectral type
B1 on the Main Sequence. Thus O--early B spectral types are going to be the
stars whose evolution we will follow in this article.

The MK system for these stars was modified and defined by Walborn (1971a).  
This was based upon spectra taken in the blue spectral region, on
photographic {\it plates}, with a (reciprocal) dispersion of 63\AA/mm. His
classification system ranged from O4 through B2.5.  Walborn (1971b) next
identified a group of early O stars near $\eta$ Carina, which extended his
spectral subtype system in a natural way to O3.  The classification for
O--early B type stars utilizes He\,{\sc i} and He\,{\sc ii} line ratios, along with
Si\,{\sc ii}, Si\,{\sc iii} and Si\,{\sc iv} features. The leading lines are as follows: He\,{\sc i}:  
$\lambda$4471 and 4387 (all units here are \AA);  He\,{\sc ii}: $\lambda$4542 and
4686; Si\,{\sc ii}: $\lambda$4128-30; Si\,{\sc iii}:  $\lambda$4552; Si\,{\sc iv} $\lambda$4089
and 4116.  Basically, the appearance of the $\lambda$4542~He\,{\sc ii} line
distinguishes the hotter O from the cooler B spectral types. The O
subtypes are defined by the ratio $\lambda$4471~He\,{\sc i}/$\lambda$4542~He\,{\sc ii}
which monotonically decreases towards earlier spectral type, such that
the $\lambda$4471 line is very weak, or totally
absent at O3.  In  luminosity V class the
$\lambda$4686~He\,{\sc ii} line behaves similarly to $\lambda$4542 although it is
a bit stronger and it persists into the earliest B type stars.

For B spectral types, $\lambda$4542~He\,{\sc ii} is absent and the He\,{\sc i} lines are
strong.  Si\,{\sc iv} lines, which are seen throughout the O types, begin to fade
at B0. Si\,{\sc iii} lines appear and the Si\,{\sc iii}/Si,{\sc iv} line ratio is useful
between B0 and B1.5.  At B2 and later subtypes, Si\,{\sc ii} appears as Si\,{\sc iii}
begins to fade and the Si\,{\sc ii}/Si\,{\sc ii} ratio comes into use.  In B and late-O
type stars, the strength of these silicon features is luminosity
dependent, being stronger in brighter stars.  The helium lines, however,
are less effected so that various silicon/helium line ratios are used as
luminosity criteria. Among late-O--early-B stars, luminosity classes of  
Ia, Ib, III and V are recognized. 

In most O type stars, though, a vastly different phenomenon begins to
become apparent in the optical: the appearance of emission lines.  
Plaskett \& Pearce (1931) noticed that in some O stars, the
$\lambda$4634-41~N\,{\sc iii} triplet came into emission along with
$\lambda$4686~He\,{\sc ii}.  These stars they labelled Of type. Walborn (1971a)  
pointed out that most O stars he observed showed the N\,{\sc iii} lines in
emission even though $\lambda$4686 remained in absorption or was not
present.  He labeled these as O((f))  and O(f) subtypes , respectively.  
We know now from detailed line modeling in O star atmospheres (e.g.,
Mihalas et al 1972) that N\,{\sc iii} emission is a result of a fortuitous
dielectronic recombination and implies nothing about the luminosity.  On
the other hand, the appearance of the $\lambda$4686~He\,{\sc ii} line, whether in
absorption, missing, or emission {\it does} say something about the
stellar wind structure (Auer \& Mihalas 1972) and the luminosity. The
status of $\lambda$4686~He\,{\sc ii} is useful as a luminosity criterion.
Following Walborn (1971a) the luminosity class of all but the latest type O
stars is as follows: if the line is in absorption, the type is O\,V; if
missing, O\,III; and if in emission, Of.  The ((f)) and (f) nomenclature is
(almost) redundant as nearly every O star, aside from the latest types,
has $\lambda$4640~N\,{\sc iii} in emission.  

%A montage of blue optical
%spectra of O stars is presented in Fig.~\ref{OB_spect}.

Recently, the O2 spectral type has been added by Walborn et al. (2002a),
such that O2--4 stars are defined by the ratio of {\it emission} lines at  
$\lambda$4058 N\,{\sc iv} and $\lambda$4634-41~N\,{\sc iii}, in contrast with the usual
He {\it absorption} line criteria for OB stars. The principal 
motivation for this was that all O stars for which 
4471 He~I was very weak or absent were previously grouped together as O3, 
yet exhibited a range of spectral morphologies, 
and stellar temperatures from quantitative analysis. The most massive,
hottest stars known, with N\,{\sc iv} $\gg$ N\,{\sc iii} emission, are re-defined as
O2 stars, and include HD\,93129A (O2~If$^{\ast}$) 
in the Carina Nebula. Until recently, this was thought to represent
one of the most massive stars in the Milky Way (Taresch et al. 1997)
but it has recently been found to be a binary (Walborn 2003).

The MK and Walborn classification systems depend on the spectra of stars
in the vicinity of the sun.  These all have a composition similar to each
other and to the sun itself. How would one classify stars that had a
substantially different abundance?  For O stars this is not a significant
problem since the spectral criteria involve hydrogen and helium lines for
the most part.  Silicon ions do play a role in classification in the late
O--early B stars and caution must be exercised in the cases of low metal
abundance.  For the LMC, with metal deficiencies of a factor 2 or 3, the
effects on the classification are barely noticeable.  For the SMC, with an
abundance down by a factor between 5 and 10, the silicon deficiency is
obvious, but the silicon ion line ratios can be used with some confidence.  
Fortunately, the luminosity of these stars is known from their membership
in the Magellanic Clouds and does not need to be established from their
spectra. Lennon (1997) provides a revision to the spectral classification
of B supergiants in the SMC.

%B stars with emission in the Balmer series (and sometimes He~I) have been
%known for a long time and have been referred to as Be stars.  This is a
%very heterogeneous spectroscopic class, encompassing main sequence stars,
%supergiants, and very young stars (Young Stellar Objects,
%YSOs).  The emission in the Main Sequence and YSO
%Be stars likely originates in a stellar disc.  In the former objects, the
%disc is a result of ejection of material from a rapidly rotating star.  
%In the latter objects, the disc arises during the stellar formation
%processes (see Chapter~\ref{chapt_birth}). Main Sequence 
%Be stars are also called ``classical
%Be stars''. In Be supergiants, the emission comes from an extended wind,
%which may also have disc-like symmetry.  A small set of O stars also have
%emission in the Balmer lines, labeled as Oe type, and can be thought of as
%analogs to the classical Be stars.

Walborn (1976) reviewed and clarified the observations that certain
O--early B stars had anomalously strong nitrogen or carbon lines, but
otherwise normal features.  These he had labelled as OBN or OBC stars. The
OBC stars were also notable for having weak nitrogen lines, but the OBN
stars have more or less normal appearing carbon features. OBN stars are
found among both main sequence stars and supergiants, while the OBC stars
are exclusive to giants and supergiants. 
Other symbol or small letter suffixes have been introduced for anomalous
OB spectra.  These will be discussed in detail in the following article. 
A review of optical classification criteria for 
O-type stars is given by van der Hucht (1996). 

In addition to optical spectral classifications, it is also possible
to assign, somewhat cruder, ultraviolet or near-IR classifications. 
Detailed atlases of 1150--1800\AA\ ultraviolet OB star 
spectra taken with the {\it IUE} satellite have been compiled by N. Walborn 
and collaborators. These have recently been supplemented with 
far-UV datasets, covering the 900--1200\AA\ region with the 
Copernicus and {\it FUSE} satellites (Walborn et al. 2002b).
Most recently of all, it has become possible to obtain classification
quality spectroscopy of OB stars in the near-IR, specifically the $K$-band
at 2$\mu$m, which was carried out in a pioneering study by
Hanson et al. (1996).
This has proved to be an extremely valuable resource by which we can
study individual hot stars in regions that are inaccessible to optical
or UV observations.

\section{Stellar temperatures}

Stellar temperatures of early-type stars, essential for subsequent
determinations of radii and luminosities, are derived from a comparison
between observed photometry or spectroscopy and models. 

% Direct - binary techniques??

\subsection{Continuum techniques using LTE models}

The simplest
case to consider is that for mid- to late- B dwarfs, with intermediate
masses, which can be
analysed using continuum energy distributions, via low dispersion 
spectroscopy or photometry. 
Str\"{o}mgren $uvby\beta$ photometry, coupled with Kurucz LTE line
blanketed model 
atmospheres, provides a powerful technique for the determination of 
$T_{\rm eff}$ in B-type and cooler stars via the Balmer discontinuity. 

Local Thermodynamic Equilibrium (LTE) 
Kurucz model atmospheres account very thoroughly for metal
line blanketing and are widely employed
for both early- and late- type stars. Although a detailed
discussion of stellar atmospheres is beyond the scope of
this book, line blanketing is the 
influence of thousands to millions of bound-bound spectral
lines on the atmospheric structure.
LTE means that the ionization state of the
gas and the populations of the atomic levels can be obtained
from the local $T_e$ and $n_e$ via the Saha-Boltzmann distribution,
 such that collisional processes occur faster than radiative processes. 

%Napiwotzki et al. (1993) have  provided a 
%discussion of the various calibrations developed over the past decade, with 
%(reddening free) Str\"{o}mgren colour-indices. 

However, in the case of early- B dwarfs, 
B supergiants and O stars,
several complications occur. Fatally for an approach based on the
Balmer jump, this feature disappears for stars with temperatures in excess
of approximately 30,000K. Although the intrinsic 
UV-optical-IR spectral  energy distribution does differ between B0
and O2--3 stars, the effect is very subtle, and masked by uncertainties
in interstellar reddening. In the case of B supergiants, the role
of the Balmer jump is weakened, given that assumed solar abundances
may no longer be appropriate.
% (See Section~\ref{composition}). 
Consequently, the temperatures of
such stars are derived via analysis of their line spectra, generally
optical photospheric lines of helium (O-type) or silicon (B stars).

\subsection{Line techniques}

Generally, radiative processes generally dominate over collisional
processes in hot star winds, so it is necessary to solve the equations
of statistical equilibrium everywhere, i.e. non-LTE. 
The problem with this is that a determination of 
populations uses rates which are functions of the radiation field, 
itself is a function of the populations. Consequently, it is necessary
to solve for the radiation field and populations simultaneously, which
is computationally demanding, and requires an numerical
iterative scheme to obtain consistency. Unfortunately, the problem
is too complex for analytical solutions.
Considerable effort has gone into developing realistic non-LTE model 
atmospheres for early-type stars in recent years by a number of
independent groups.  Unfortunately, consistently 
treating metal `line blanketing' in extended non-LTE atmospheres
is computationally demanding.  Consequently, the earliest attempts
by D. Mihalas in the late 1960's suffered from simplifications that
only became surmountable two decades later, at which time large
amounts of atomic data became available from the Opacity Project 
which was led by M. Seaton.

\begin{figure}
\includegraphics[height=7.3cm]{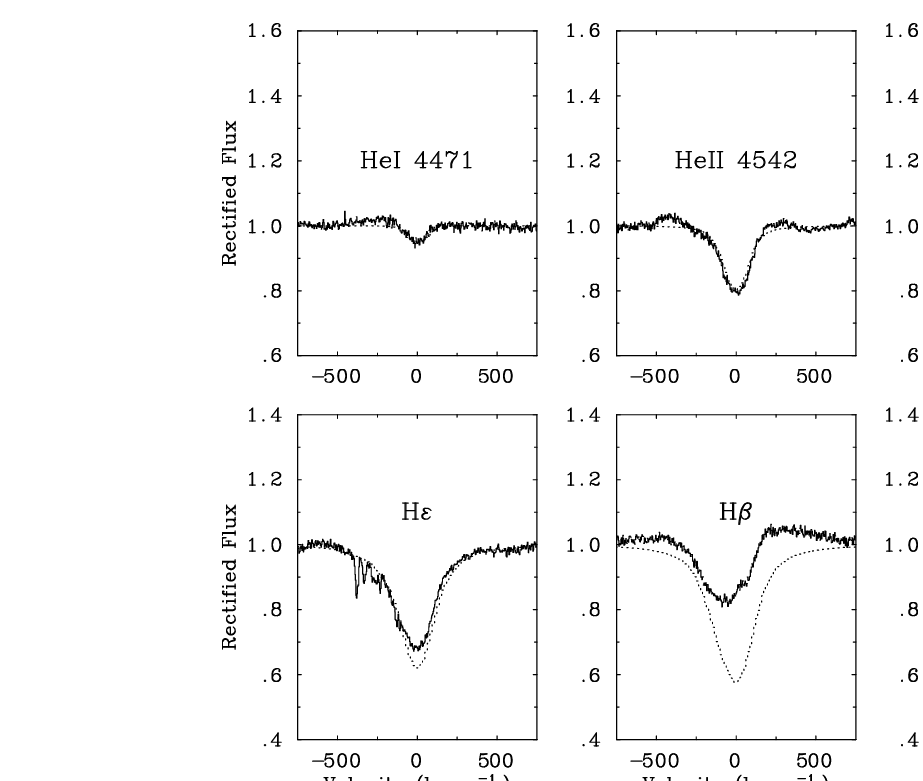}
%\qquad
\caption{Comparison between optical
line profiles of HDE\,269698 (O4Iaf+)
and synthetic spectra (dotted) calculated
using an unblanketed, plane-parallel version
of {\sc tlusty} with \Teff=46.5kK, $\log L/L_{\odot}$=6.25,
$\log g$=3.7 (Crowther et al. 2002)}
%\end{figure}
%\begin{figure}[b!]
\includegraphics[height=7.3cm]{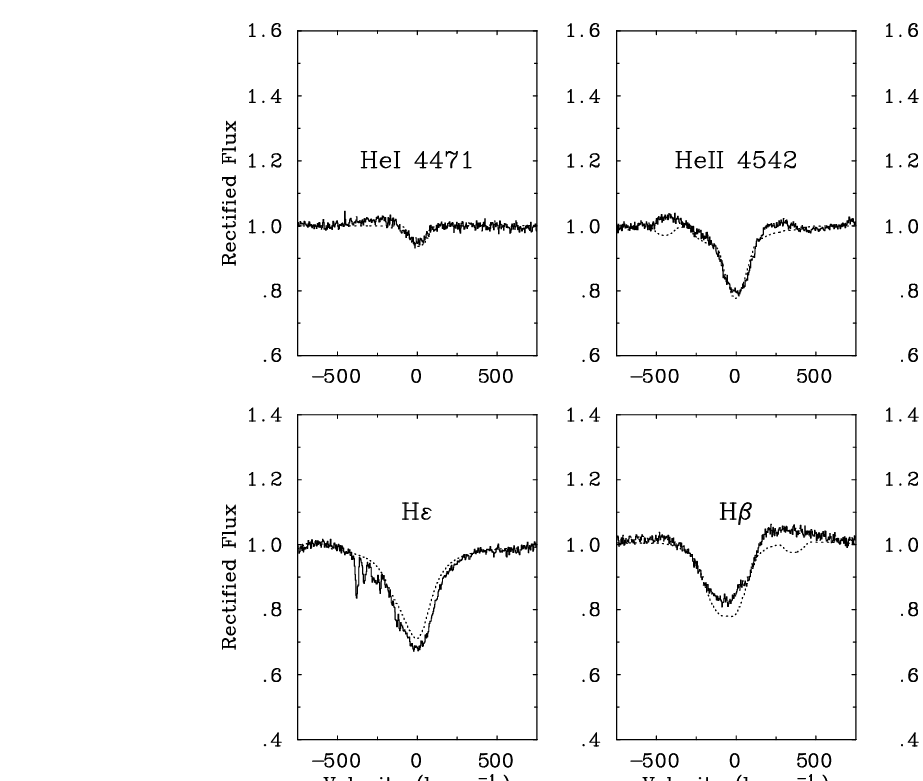}
%\qquad
\caption{Comparison between optical
line profiles of HDE\,269698 (O4Iaf+)
and synthetic spectra (dotted) calculated
using the line blanketed, spherical code
{\sc cmfgen} with \Teff=40kK, $\log L/L_{\odot}$=5.98,
$\log g$=3.6, \Md=8.5$\times 10^{-6} M_{\odot}$ yr$^{-1}$,
$\beta=1$, \vinfty=1750 \kms (Crowther et al. 2002)}
\end{figure}

From such techniques, temperatures of hot stars
are obtained using ionization equilibrium techniques, i.e.
from suitable diagnostic lines of adjacent ionization stages. 
Naturally, determinations of \Teff\ from model atmospheres are
critically dependent on assumptions going into those models
and the accuracy with which surface gravities and elemental abundance
 ratios are determined. 
The majority of detailed analyses of OB stars have been subject to the 
assumption of hydrostatic equilibrium in a so-called 
`plane-parallel' approach  (e.g.
Herrero et al. 1992; Puls et al. 1996), where the spherical extension of
the star is neglected. 
Optical He\,{\sc i-ii} lines generally provide temperature diagnostics for
O stars (e.g. Herrero et al. 1992) while Si\,{\sc ii-iv} lines
are used for B stars (e.g. McErlean et al., 1999). The typical range of
temperatures spanned by O stars is \Teff~$\simeq$50,000K 
at the earliest subtypes
to \Teff~$\simeq$30,000K at 
B0~V, and \Teff~$\simeq$10,000K at A0~V. 

The temperature scale of OB dwarfs and supergiants 
has been compiled by various authors, most recently
by Vacca et al. (1996). However, a word of caution is necessary
at this point.
This, and all previous compilations 
 are based on plane-parallel pure H-He models. The effect of line 
blanketing from millions of metallic lines on the emergent spectra
and ionization balance was neglected in most literature papers. 

Within the past few years several non-LTE model atmosphere codes
allowing for metal line blanketing have been developed. {\sc tlusty}
represents the current state-of-the-art for `plane-parallel' studies
of OB dwarfs  (Hubeny \& Lanz 2003), whilst spherically extended
non-LTE line blanketed codes include {\sc cmfgen} (Hillier \& Miller 1998),
{\sc WM-basic} (Pauldrach et al. 2001), {\sc fastwind} 
(Santaloya-Rey et al. 1997)
and the Potsdam code (Gr\"{a}fener et al 2002). Recent studies of OB dwarfs  
have been carried out using {\sc cmfgen} and {\sc tlusty}, 
employing metal line blanketing,
resulting in systematically lower temperatures 
(e.g. Martins et al. 2002; Bouret et al. 2003). 

The case for O supergiants
is further affected by the presence of strong stellar winds. These
may contaminate (`fill-in') photospheric absorption lines, 
causing temperatures to be overestimated still further in the 
standard plane-parallel assumption (Schaerer \& Schmutz
1994). Bohannan et al. (1990)
demonstrated this for the O supergiant, $\zeta$ Pup (O4\,If) in 
which wind blanketing reduced previous determinations by $>$10\%. 
Putting
stellar winds and line blanketing together can cause downward revisions in
\Teff\ by up to 20\% in extreme OB supergiants (Crowther et al. 2002;
Herrero et al. 2002). This is illustrated in Figures 1 and 2, which
show respectively, spectral fits to HDE\,269698 (O4\,Iaf+) in the LMC using
a unblanketed plane-parallel code, revealing \Teff=46.5kK, in which the 
wind lines are neglected, and blanketed, spherically extended code, 
revealing \Teff=40kK plus simultaneous fits to wind features 
in the optical, UV and far-UV (Crowther et al. 2002).
%
%In the case of $\zeta$ Pup, early non-LTE models
%suggested a stellar temperature of 48kK, whilst wind blanketing reduced
%its temperature to 42kK and recent results based on 
%line blanketed, spherically extended codes  indicate values as low as
%39kK (Crowther et al. 2002; Gr\"{a}fener, 
%priv. comm.) 
Such major downward revisions have a strong 
influence on the stellar luminosity of individual stars, which of course
scale as $T_{\rm eff}^4$.

\section{Masses}

The mass of a star is perhaps its most fundamental property, yet it
is generally very difficult to constrain for massive stars to better than
a factor of two. This is because young massive stars of differing ages
and masses cluster together very tightly in the upper left H-R diagram.
There represent two, model dependent,
methods that may be applied to derive a stellar mass, plus a third,
more direct technique 
if the star happens to be a member of a double-lined, short period binary. 

\begin{figure}
\includegraphics[width=10cm]{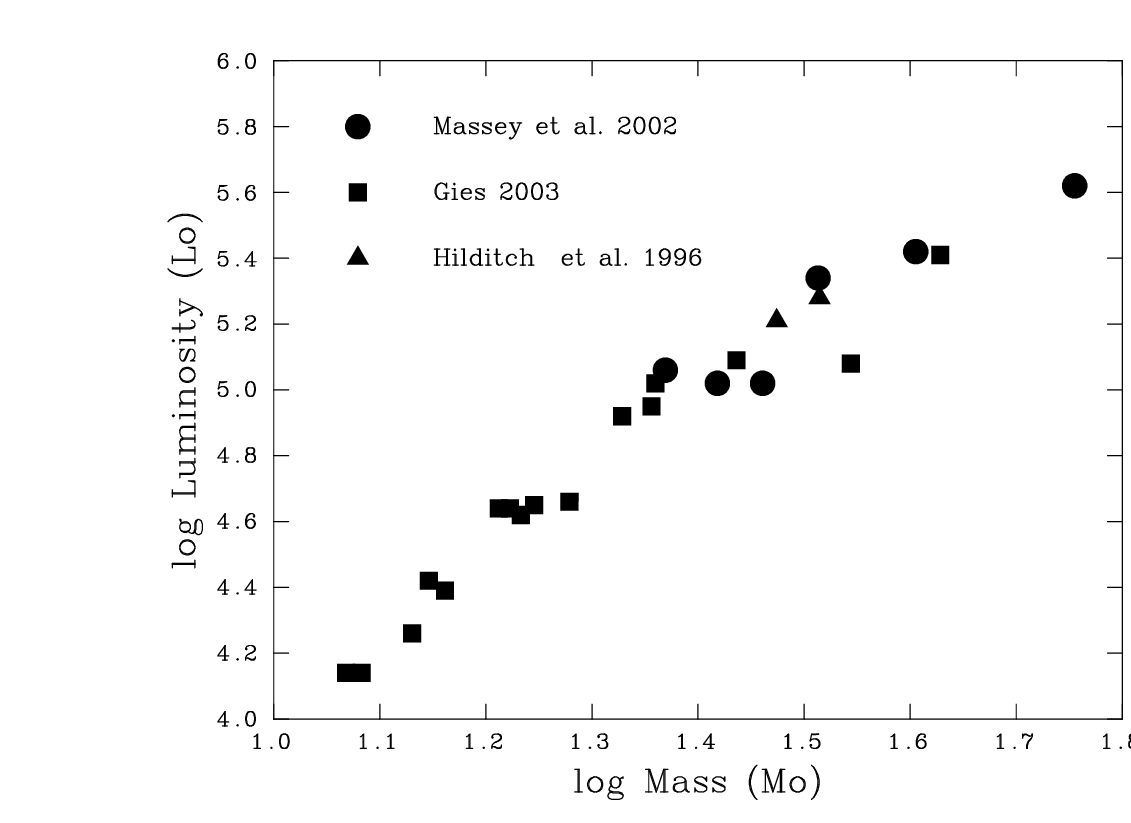}
%\qquad
\caption{Mass-luminosity relationship for young OB stars
from binary orbits, taken from the Gies (2003) compilation,
Hilditch et al. (1996) for DH Cep, plus Massey et al. (2002) for
LMC stars}
\end{figure}

In general, massive stars are not members of double-lined, short
period binaries. The primary 
method of deriving a star's mass is via a determination of
the surface gravity from fitting line profile 
wings of Balmer hydrogen  
absorption lines. The spectroscopic mass is determined from this
gravity determination, using 
\begin{equation}
g = \frac{GM}{R^2}
\end{equation}
In reality, an `effective gravity' is determined for hot
stars due to the complication of radiation pressure, as will be
discussed later in this article, and is expressed as $\log g$ in cgs units.
Typical values for OB dwarfs are $\log g=$4--4.5, with somewhat
lower gravities, implying more extended atmospheres, for supergiants,
e.g. $\log g\sim$3 in late O or early B supergiants. 
 High dispersion optical spectroscopy is needed in order
to model the Balmer line wings accurately enough.

Alternatively, the position of a star in the H-R diagram can be 
compared to predictions from evolutionary tracks to extract a stellar
mass. This is most straightforward for main sequence stars, as
it is often uncertain whether individual OB supergiants are evolving
to cooler temperatures immediately after the main sequence, 
or blueward after an excursion to the red part of the H-R diagram.
Surface abundances will generally allow one to discriminate between these
alternatives, but these are difficult to reliably measure for hot stars.

Herrero et al. (1992) first established a ``mass discrepancy'' between the
sppectroscopic mass determinations and those resulting from comparison
with evolutionary tracks. This remains to this day, with evolutionary
masses up to a factor of two times higher than spectroscopic masses for OB
stars. Potential problems
lie in both the stellar luminosities obtained from model analyses, see above,
and in the evolutionary models. For the latter,  there was previously
a rather tight correlation between luminosity and mass for hot stars, but
the recent inclusion of rotation into evolutionary calculations has 
revealed that alternative initial rotational velocities can lead to a 
considerable spread in (\Teff, $\log L)$ for stars of identical 
initial mass.

The most accurate method of determining masses for OB stars, and so
establishing which, if any, of these two approaches is correct, 
is if these are members of binary systems. Unfortunately, very few
double-lined O+O systems have sufficiently short orbits to permit
spectroscopic monitoring and determination of complete light curves
practical. 
Gies (2003) presents a summary of recent studies of detached (non-interacting)
 massive binaries. These permit a direct mass-luminosity 
relation for young OB stars, as shown in Fig.~3. Between $10-30 M_{\odot}$,
$L \propto M^{2.5}$, above which $L \propto M^{1.75}$.
%, in which recent results
%from Massey et al. (2002) for massive binaries in R136 in the LMC are
%included.
%
%FIGURE
%For those that have been studied, mass-transfer may 
%have occured, hindering the reliability of
%evolutionary models for single stars. This mechanism
%will be considered in Chapter~\ref{chapt_binary}.
% ideally eclipsing so that the orbital
%inclination is well constrained 
%Recent attempts at shedding light on this discrepancy are rather inconclusive.
%Penny et al. (1999)  found masses for 25$M_{\odot}$ for
%each component of HD~152248 (O7I+O7I) 
%25$M_{\odot}$, significantly lower than 35--40$M_{\odot}$ from
%evolutionary models for single stars. It is probable that allowance
%for rotation in evolutionary models will improve consistency for low
%gravity giants and supergiants (Meynet \& Maeder 2000). Such difficulties
%should not be present for high gravity dwarfs. 
%Indeed,

Massey et al. (2002) have used very massive binaries in the 
LMC R136 cluster to confirm evolutionary mass determinations, 
including a mass of 57$M_{\odot}$ for the O3~V primary of R136 \#38
in close agreement with 53$M_{\odot}$ from evolutionary models. Nevertheless,
problems do remain in some cases.
% e.g. the primary in the
%Carina O3~V+O8~V binary HD~93205 does appear to be less massive than
%predictions would suggest (Morrell et al. 2001).
Consequently, the jury remains out for the present on this question.
An added complication with the binary method is that stellar temperatures
and bolometric corrections need to be derived or adopted from
calibrations. Errors in such calibrations may exacerbate differences.
Ultimately, the use of modern temperature calibrations in binary studies
and line blanketed stellar atmosphere models will likely resolve evolutionary
and spectroscopic masses for early-type stars (Herrero et al. 2002).

\section{Stellar Wind Properties}

Winds are ubiquitous amongst massive stars, although the physical processes
involved depend 
upon the location of the star within the H-R diagram. 
Mass-loss crucially affects the evolution and fate of a 
massive star, while
the momentum and energy expelled contribute to the dynamics and energetics
of the ISM.
% The interested reader is referred to
%the monograph by Lamers \& Cassinelli (1999) on the topic of stellar winds,
%or Kudritzi \& Puls (2000) for a more detailed discussion of mass-loss from
%OB stars. A summary of typical wind velocities and mass-loss rates 
%is presented in Table~\ref{wind_properties}

The existence of winds in O-type stars has been established 
since the 1960's, when the first rocket UV observations 
revealed the characteristic P~Cygni signatures of 
mass-loss from \civ, \siiv, and \nv\ (Morton 1967). A theoretical
framework for mass-loss in hot stars was developed by Castor, Abbott
\& Klein (1975), known as CAK theory via line-driven radiation pressure. 
The fundamental characteristics of stellar
winds are velocities and mass-loss rates. The former can be directly
observed, whilst the latter relies of varying
complexity of theoretical interpretation.

\subsection{Theory of radiatively driven winds}

The basic mechanism by which hot star winds are driven is the transfer
of photospheric photon momentum to the stellar atmosphere through
absorption by spectral lines. The velocity reached
far from the star tends towards a terminal value, $v_{\infty}$, due to
the geometrical dilution of the photospheric radiation field, whilst
the mass-loss rate relates to the velocity field, $v(r)$ and density,
$\rho(r)$ via the equation of continuity 
\begin{equation}
\dot{M} = 4 \pi r^2 \rho(r) v(r)
\end{equation}
for a spherical, stationary wind.

%\subsection{Metallicity dependence}

The properties of stellar winds depend on both the number of metal
lines available
to absorb photon momentum, and on their ability to absorb, i.e. their
optical thickness. Analytical solutions involve force multiplier
parameters, $\alpha$,
$\delta$ and $k$ which characterize the outflow (Kudritzki et al.
1989). Of these, $\alpha$
controls the fraction of optically thick/thin
lines, $k$ the number of strong lines, 
and $\delta$ the ionization balance.
%The radiative acceleration, $g_{\rm line}$
%can be written in terms of these force multipliers as follows,
%\begin{equation}
%g_{\rm line} \propto 
%k \left(\frac{\sigma_{e}v_{\rm th}\rho}{\dv}{dr}\right)^{-\alpha}
%  \left(\frac{n_e}{W(r)}\right)^{\delta}
%\end{equation}
%where $\sigma_e$ is the electron scattering cross-section,
%$v_{\rm th}$ is the mean thermal velocity, 
%$n_e$ is the electron density, and $W(r)$ is the dilution factor. 

It can be shown that, 
\begin{equation}
v_{\infty} \propto \left(\frac{\alpha}{1-\alpha}\right)  v_{\rm esc}
\end{equation}
and
\begin{equation}\label{kappa}
\dot{M} \propto k^{1/(\alpha-\delta)}
\end{equation}

From above, there is expected to be a close relationship between
the terminal velocity and the escape velocity,
\begin{equation}\label{vesc}
v_{\rm esc} = (2 g_{\rm eff} R)^{0.5}
\end{equation}
where the effective gravity  
\begin{equation}\label{geff}
g_{\rm eff} = \frac{GM}{R^2}\left(1 - \Gamma\right)
\end{equation}
is the stellar gravity corrected for the reducing effect of 
radiation pressure (via Thompson scattering) 
on the gravitational potential 
via $\Gamma$ which is related to the stellar luminosity and mass.
It is clear that the determination of stellar
escape velocities depends on knowledge of stellar masses and radii.

Recent theoretical studies have indicated that
CNO elements are principal line drivers for the outer, supersonic part
of the wind, whilst iron group elements are responsible for the
inner, subsonic part (Vink et al. 1999). The former determine the wind
velocity, and the latter the mass-loss rate, at least for compositions
close to Solar metallicity.

Hot stellar winds are predicted to depend on the metal content, $Z$,
as follows.  Puls et al. (1996) showed that $k \propto Z^{1-\alpha}$, such 
that Eqn~\ref{kappa} implies 
\begin{equation}
\dot{M} \propto Z^{(1-\alpha)/(\alpha-\delta)}
\end{equation}
Inserting typical values of $\alpha$(=2/3) and $\delta$(=1/10) suggests
an exponent of $\sim$0.6, close to 0.5 
deduced by Kudritzki et al. (1987) and 0.7 derived by Vink et al. (2001).
Since $\alpha$ is also predicted to be a function of metallicity, 
Leitherer et al. (1992) derived $v_{\infty} \propto Z^{0.13}$ for OB stars.  

Puls et al. (1996) and Kudritzki et al. (1999) obtained
a 
theoretical
relationship between the bolometric luminosity of an early-type star
and the wind momentum, $\dot{M} v_{\infty}$ modified by the square-root
of the stellar radius, i.e. the so-called wind-momentum relation, 
as follows:

\begin{equation}
\dot{M} v_{\infty} (R/R_{\odot})^{0.5} \propto L^{1/(\alpha-\delta)}
\end{equation}

For O stars, taking $\alpha=0.6, \delta=0.05$, implies a luminosity
exponent of $\sim$1.8, close to that observed for Galactic 
O supergiants ($\sim$1.5) and dwarfs ($\sim$1.6) 
according to Kudritzki \& Puls (2000). 
The exponent for B supergiants is different, due to the change in 
ionization of the elements contributing to the radiative line acceleration.
Differences in coefficient follow from the recent  revision in temperature
calibration for O and early B stars (Herrero et al. 2002).
This relationship provides a potential independent method of determining
distances to galaxies beyond the Local Group, although it is primarily
of use for (less luminous) A supergiants which are visually the brightest
hot stars in galaxies.
 
\subsection{Wind velocities}

Ultraviolet P~Cygni profiles, ubiquitous in O-type stars 
 provide a direct indication of stellar winds.
%Figure~\ref{P_Cygni} illustrates the formation of a P~Cygni line profile.
Wind material approaching the observer within a column in front of
the star is blueshifted by the Doppler effect. Scattering of radiation
out of this direction causes a reduction in the observed flux on the
blue side of the profile. From the regions on either side of this column
wind material may scatter radiation towards the observer. This may occur
from either the approaching (blue-shifted) or receding (red-shifted)
hemisphere, the extra flux seen by the observer, originating 
from this scattered radiation, is a symmetric emission component on both
sides of the profile. The overall effect is therefore, asymmetric
with blue absorption and red emission. The wavelength of the blue edge
of the absorption provides a measure of the asymptotic wind velocity. 
Accurate wind velocities of OB and W-R stars can
be readily obtained in this way from {\it HST} or {\it IUE} observations of saturated 
\civ, or \siiv\ P~Cygni profiles
(Prinja et al. 1990; Lamers et al. 1995). The range in velocities
encountered is considerable: $v_{\infty}$=3500 km\,s$^{-1}$,
corresponding to $\geq$1\% of the speed of light, in the
earliest O stars, to 100 km\,s$^{-1}$ in some AB supergiants.

The number of OB stars observed in our Galaxy, principally with {\it IUE},
far exceeds that from external galaxies due to their relative UV brightness.
Nevertheless, sufficient extragalactic hot stars have now been observed,
principally with {\it HST}, for us to be able to make useful comparisons. 
Wind velocities of LMC
O stars differ little from Galactic counterparts (Garmany \& Conti 1985), 
while a more prominent effect {\it is} observed in the SMC, 
particularly amongst early O stars (Walborn et al. 1995;
Prinja \& Crowther 1998).

Theoretically, wind velocities of hot stars are expected to follow
a so-called $\beta$-law,
 
\begin{equation}
v(r) = v_{\infty} \left( 1 - \frac{R}{r} \right)^{\beta}
\end{equation}

where $R$ is the stellar radius and $\beta\sim$0.8--1.5. 
UV line profile modelling appears to confirm expectations. %(Groenwegen et al. 1989).

In spite of often uncertain stellar masses, Lamers et al. highlighted
the so-called `bi-stability' jump in wind properties around B\,1 
(\Teff $\sim$21,000K), 
above which  $v_{\infty} \simeq 2.65 v_{\rm esc}$, 
and $v_{\infty} \simeq 1.4 v_{\rm esc}$. This is thought to 
result from the change in ionization of the 
dominant elements contributing to the line force (Vink et al. 1999).

\subsection{Mass-loss rates}

 Observationally, estimates of OB mass-loss rates are
obtained
from radio continuum fluxes, optical/IR line profiles, or UV wind lines.
In extreme cases, $\dot{M}$ may exceed 10$^{-5}$ $M_{\odot}$ yr$^{-1}$
in some early O supergiants, but values 10-100 times lower are more
typical. 

\subsubsection{IR--radio continua}

Winds in hot stars can be readily observed at IR-mm-radio wavelengths
via the free-free (Bremsstrahlung) `thermal' excess caused by the 
stellar wind, under the assumption of homogeneity and spherical symmetry.
Mass-loss rates can be determined via application of relatively simple
analytical relations (e.g. Wright \& Barlow 1975) which
reveal that the continuum flux, $S_{\nu} \propto \nu^{0.6}$,  in a spherically
symmetric, isothermal envelope which is expanding at constant velocity. 
The emergent flux is set by the distance to the 
star $d$, mass-loss rate and terminal velocity as follows

\begin{equation}
S_{\nu} \propto \left( \frac{\dot{M}}{v_{\infty}}\right)^{4/3} 
\frac{\nu^{0.6}}{d^2}
\end{equation}

Other factors, including differences in 
composition, ionization balance, electron temperature
generally play a minor role in normal OB stars. 
Barlow \& Cohen (1977) used this approach
together with IR excess fluxes to determine 
mass-loss rates for a large sample of 
Galactic OBA supergiants. Radio measurements of OB stars were carried
out by Abbott and collaborators with the VLA, 
as summarised by Bieging et al. (1989), and by Leitherer and collaborators
with the Australia Telescope (Leitherer et al. 1995). 
Unfortunately, OB stars with relatively weak winds do not show
a strong IR excess or radio flux,  so mass-loss results from this
technique have solely been for nearby hot stars with dense winds. 

Collisions between 
stellar winds from stars in a binary system will 
cause non-thermal (synchrotron)
radio emission, so care needs to be taken against overestimating
mass-loss rates from (apparently) single stars in this way.
Generally this is accomplished by carrying out 
observations at multiple radio frequencies.

\subsubsection{Optical and IR line profiles}

H$\alpha$ has long been recognized as the prime
source of mass-loss information in early-type stars. However,
accurate determinations rely upon a complex treatment of the lines and
continua, i.e. non-LTE, spherically extended 
models which treat the sub- and supersonic atmospheric structure.
Such techniques have already been touched upon with regard to
surface temperatures of hot stars.
Problems with using fits to H$\alpha$ to derive mass-loss
rates may include blending with He\,{\sc ii} $\lambda$6560 
in O stars, together with
nebular contamination from H\,{\sc ii} regions, 
and broadening of the central emission component by stellar rotation.

Hot stars in the Magellanic Clouds provide us with the means to compare
empirical mass-loss properties with predictions at lower $Z$. 
Although the number of stars analysed for the Magellanic Clouds
remains small (Puls et al. 1996), the mass-loss rates of LMC and SMC 
O stars do appear to be lower than Galactic counterparts, in reasonable
accord with theoretical predictions.

Although moderate to high spectral resolution observations of OB stars
are rare at near-IR wavelengths, there exist wind diagnostics
analogous lines to optical lines 
(Bohannan \& Crowther 1999). This provides the prospect of
the determination of mass-loss properties for hot stars obscured at 
optical wavelengths via IR diagnostics.
%This will be discussed in detail in the following article.

\subsubsection{UV P~Cygni profiles}

UV P~Cygni profiles from metal resonance lines 
provide direct evidence for mass-loss in hot stars.
Unfortunately, mass-loss determinations require knowledge of
elemental abundances, the degree of ionization of the
ion producing the line, and knowledge about the form of the
velocity law. 

Recent analyses of hot star winds via UV spectroscopy are generally
carried out using the Sobolev with Exact Integration (SEI) method  
(Lamers et al. 1987; Haser et al. 1998), named after the pioneer
of stellar wind studies, V. Sovolev from the late 1950's.
In the SEI approach, UV P~Cygni
line profiles are fitted to reveal $\dot{M} q$, where $q$
is the, {\it a priori} unknown, 
fractional population of a particular ionization
stage within a particular element. Direct ionization 
information is limited to observed ions from {\it IUE} or {\it HST} spectroscopy:
\nv, \siiv, and \civ, so mass-loss rates may only be estimated via
ionization balance information, generally from non-LTE atmospheric models. 
%In view of these difficulties, 
%Lamers et al. (1999) have combined column densities from unsaturated UV
%5P Cygni profiles with independently determined mass-loss rates (from
%radio or H$\alpha$, see above) to derive empirical ionization fractions. 
%A more direct approach is now available for hot stars by incorporating
Far-UV {\it FUSE} observations complement previous observations, providing
information on many more ionization stages, including \niii, \ciii, 
\siv, \svi, \pv. Massa et al. (2003) have used SEI modelling of 
line profiles measured {\it FUSE} spectroscopy of LMC  OB stars to derive empirical
ionization balances, albeit relying on  
theoretical mass-loss rates from Vink et a. (2001).

%more ionization stages, e.g. C~III, N~III, O~VI,
%S~IV, VI, P~IV-V. SEI modelling of FUSE observations of Magellanic Cloud
%early-type stars is currently underway (e.g. Massa et al. 2003).

\subsection{Structure}

There is extensive observational evidence that hot star winds are not
smooth, steady outflows. Intensive monitoring of unsaturated UV P~Cygni
lines with {\it IUE} led to extensive variability in the absorption components.
To produce such variability, the wind structure must be on a large
scale, covering a substantial fraction of the disk. In contrast, emission
components of P~Cygni profiles were found to be rather less variable, 
since it forms from a more global average of the wind. Subsequent optical
studies have also led to absorption and emission line variability.

There is also indirect evidence that OB winds have a turbulent structure.
Saturated P~Cygni profiles have extended black troughs, which are thought
to be a signature of a highly non-monotonic wind. Soft X-ray emission is
also observed, which is thought to originate from embedded wind shocks
(Chlebowski et al. 1989; Lucy \& White 1980). Such 
shocks are also thought to be the means by which high ionization
stages are observed, most notably \nv, in stars for which the \Teff\ is too
low from normal means. This is via a process known as 
super- (or Auger) ionization.

A likely explanation for this turbulent structure is theoretical evidence
for a strong instability of line driving to small-scale velocity perturbations.
There is a strong potential in line scattering to drive 
wind material with accelerations that 
greatly exceed the mean outward acceleration. 
Simulations demonstrate that this instability may lead
naturally to a highly structured flow dominated by 
multiple shock compressions (e.g. Runacres  \& Owocki 2002). 
%Nevertheless, it remains to be demonstrated whether simulations are able
%to explain observational features, such as the soft X-ray emission.
Such predictions are broadly consistent with recent {\it Chandra} X-ray
line spectroscopy of $\zeta$ Pup (Cassinelli et al. 2001) 
and other early O stars.
Interpretation of the X-ray spectroscopy of $\theta^1$ C Ori indicates
a magnetic origin.
  
If hot star winds are structured, or clumped, it remains to be seen whether
derived mass-loss rates of
OB stars, as derived from optical or radio measurements, 
represent over-estimates of true values. 
Overall consistency between e.g. H$\alpha$
and radio determinations (Lamers \& Leitherer 1993) is rather good, suggesting
that if winds are highly clumped, they are similar on scales of $\leq 1.5
R_{\ast}$  (H$\alpha$) and $\sim$1000 $R_{\ast}$ (radio). 
Time-averaged line profiles of OB stars that have been intensively monitored
do remain remarkably stable on scales of years. Nevertheless, recent 
{\it FUSE} spectroscopic observations of O supergiants in the Magellanic Clouds
require that their winds are clumpy in order to simultaneously reproduce
far-UV and optical wind lines, principally \pv, unless
phosphorus is underabundant relative to other elements in these galaxies
(Crowther et al 2002). Evans et al. (2003) arrive at similar conclusions
in late O and early B supergiants using \siv. Although the ISM abundance
of Phosphorus in the Magellanic Clouds is uncertain, Sulphur is well 
constrained. Consequently,  the most credible possibility is that 
OB supergiants are clumped (see also Massa et al. 2003).

%In contrast, evidence for highly clumped winds for W-R 
%stars appears overwhelming. 
%Line profiles show small-scale structures or blobs (Lepine et al. 2000),
%as shown in Fig.~\ref{hd192103_blobs}, 
%ionized nebulae are highly clumped (Grosdidier et al. 1998), and individual
%spectral lines, formed at $\sim 10 R_{\ast}$ can be used to 
%estimate (small) volume filling factors (Hillier 1991).

%Since the region where H$\alpha$ forms in OB stars 
%($\leq 1.5 R_{\ast}$) is not thought to be heavily structured, mass-loss 
%determinations may not be greatly affected by clumping. 

\section{Rotation}

During the formation process of all stars, including hot luminous OB stars,
the conservation of angular momentum implies that they will commence
their Main Sequence 
life rotating very fast, perhaps close to break-up velocity. 
Due to rotational broadening mechanisms, we
are able to measure current OB rotational velocities from optical or UV
photospheric line profiles. For OB stars, the principal compilations are
by Conti \& Ebbets (1977) and  Howarth et al. (1997). 

Due to projection effects,
$v \sin i$ is actually 
measured, where $i$ is the inclination of the system on
the sky as viewed by the observer. In addition, measured projected
rotation velocities underestimate the true projected equatorial 
velocity, $v_{e} \sin i$,
due to limb darkening in rapid rotators (Collins \& Truax 1995).
From a sample of over 400 OB stars,
Howarth et al. determined a median value
of $v \sin i$ = 90 km\,s$^{-1}$, which de-projects to a value 10\% 
higher on average. The determination of rotational velocities 
is aided by some very slow rotators which serve as useful templates.
 $\tau$ Sco (B0.2~IV) is the most famous example, 
with $v \sin i<$ 5km\,s$^{-1}$. 

One finds that the O 
Main Sequence distribution is much broader than that of supergiants, with
only 3 supergiants amongst 33 stars with $v_{r} \sin i > 200$ km\,s$^{-1}$.
This can be understood in that as a star evolves from the ZAMS, its
rotational velocity will decrease due to conservation of angular momentum,
lowering the limit to the distribution. The various distributions are
summarised by Howarth et al. (1997).

Theoretically, Friend \& Abbott (1986) first considered the effect of
centrifugal acceleration on hot star winds, extended by Bjorkman \& 
Cassinelli (1993) to allow for the azimuthal dependence, who found
that $\dot{M}$ increases towards the equator, and $v_{\infty}$ decreases,
leading to an oblate `wind-compressed disk'. This has been widely applied
to Be stars, a subset of B stars exhibiting emission lines.
Subsequently, it was established by Owocki et al, (1996) 
that a quite different geometry may result, 
by additionally allowing for non-radial  components 
of the line force, and the increased polar radiation flux 
via gravity darkening (von Zeipel 1924). Due to a reduced escape velocity, 
the 
equatorial velocity law is slower than at the pole, inhibiting the 
{\it equatorial} 
disk and the original scaling of mass-loss can be reversed, such 
that a {\it prolate} geometry results. 
By way of example, an early-B dwarf rotating 
at 85\% of its critical, or break-up, 
velocity was found by Petrenz \& Puls (2000) 
to have a strongly prolate structure ($\rho_{\rm pole}/\rho_{\rm eq}
\leq$15) with a polar (equatorial) terminal velocity of 1030 (730)
km\,s$^{-1}$. The global mass-loss rate was not found to deviate from
its 1D value by greater than 10--20\%, except for supergiants close 
to the Eddington limit.

\section{Composition}\label{composition}

The French philosopher Auguste Comte famously, and inaccurately,
predicted of stars in 1835 that 
{\it `.. we would never know how to study by any means their 
chemical composition ..'} 
Fortunately, theoretical knowledge
of stellar atmospheres, atomic data, together with high-resolution 
spectrograph's now
permits abundance determinations for most normal stars,
including hot, luminous OB stars. 

This can only readily be carried out using optical photospheric lines of
H, He and metals, based on 
the same non-LTE techniques described above that are used to derive 
\Teff, $\log g$. Consequently, it is generally
solely hydrogen-helium abundances which have been determined
to date for a relatively
large number of O stars (e.g. Herrero et al. 1992). 
Whilst most main sequence
stars showed normal He abundances, He/H$\approx$0.1 by number, 
as expected, rapid rotators were surprisingly found 
to have enriched He at their surfaces,  suggesting that rotational
mixing has brought products of core nucleosynthesis to the surface while
the star is still on the main sequence. Advances in model atmosphere
modelling since then have generally supported these results.
Recently, Bouret et al. (2003) have analysed the CNO abundances of 
a sample of SMC O dwarfs using non-LTE line blanketed models. 

Studies of O and early-B supergiants reveal modest enrichment of He, whilst 
CNO elemental abundances have mostly been lacking until recently.
Smith \& Howarth (1994) studied late OC, O and ON supergiants, 
revealing
increasing helium abundances for these, suggesting an evolutionary sequence
in support of the morphological criteria. 
However, great care should be taken when analysing OB stars for abundance
determinations since neglect of `micro-turbulent' line broadening may imply
erroneous He-enrichments (Smith \& Howarth 1998; McErlean et al. 1998).
There is clear evidence for enhanced nitrogen, accompanied with decreased
carbon in OB supergiants from studies by Crowther et al. (2002),  Hillier
et al. (2003) and Evans et al. (2003). Notably, an ON9.7~Ia+ star in
the LMC was found to be significantly more chemically processed than
an otherwise identical O9.7~Ia+ star, whilst OC and BC stars were
confirmed as CNO-`normal'. CNO abundances in OB supergiants in the SMC 
were found to exhibit similar degrees of chemical processing as A supergiants
studied by Venn (1999).

%, O-type) and
%McErlean et al. (1999,l B-type) using optical and/or 
%UV line profile modelling. Crowther et al. 

\section{Further reading}

Kudritzki \& Puls (2000) have provided an excellent Annual Review 
 article on the properties of OB stars, whilst the book by
 Lamers \& Cassinelli (1999) discusses Stellar Winds of both 
early- and late- type stars. Recent IAU Symposia
involving massive stars
include \#212 (van der Hucht et al. 2003) and \#215  (Maeder \& Eenens 2003).
%, Stellar Astrophysics for
%the Local Group (1998).

%\acknowledgements
%Thanks to Peter Conti for his contribution to this work.

%\section{CNO abuindances}

%%-----------------------------
%%      your bibliography
%%-----------------------------


\begin{thebibliography}{99}

\bibitem[]{} Auer L.H. \& Mihalas D. 1972, ApJS 24, 193

\bibitem[]{} Barlow M.J. \& Cohen M., 1977, ApJ 213, 737

\bibitem[]{} Bieging J.H., Abbott D.C. \& Churchwell E.B., 1989, ApJ 340, 518

\bibitem[]{} Bjorkman J.E. \& Cassinelli J.P., 1993 ApJ 409, 429

\bibitem[]{} Bohannan B. \& Crowther P.A., 1999 ApJ 511, 374

\bibitem[]{} Bohannan B., Abbott D.C., Voels S.A. \& Hummer D.G., 1990, ApJ 365, 729

\bibitem[]{BLM} Bouchet P., Lequeux J., Maurice E., Prevot L. \& Prevot-Burnichon M.L., 1985, A\&A 149, 330

\bibitem[]{} Bouret J.C. Lanz T., Hillier D.J. et al. 2003, ApJ in press (astro-ph/0301454)

\bibitem[]{} Cassinelli J.P., Miller N.A., Waldron W.L., 
MacFarlane J.J. \& Cohen D.H., 2001, ApJ 554, L55

\bibitem[]{} Castor J.I., Abbott D.C. \& Klein R.I., 1975, ApJ 195, 157

\bibitem[]{} Collins G.W. II \& Truax, R.J., 1995, ApJ 439, 860

%\bibitem[]{} Conti P.S., Leep M.E., Perry D.N., 1983, ApJ 268, 228

\bibitem[]{} Conti P.S. \& Ebbets D., 1977, ApJ 213, 483

\bibitem[Chlebowski Harnden \& Sciortino 1989]{1989ApJ...341..427C} 
Chlebowski, T., Harnden, F. R. , Jr. \& Sciortino, S. 1989, ApJ, 341, 427 

\bibitem[]{} Crowther P.A. \& Dessart L., 1998, MNRAS 296, 622

\bibitem[]{} Crowther P.A., Hillier D.J., Evans C.J. et al. 2002, ApJ 579, 774

\bibitem[]{} Evans C.J., Crowther P.A. et al. 2003, MNRAS in preparation

\bibitem[]{} Friend D.B. \& Abbott D.C., 1986, ApJ 311, 701

\bibitem[]{} Garmany C.D. \& Conti P.S., 1985, ApJ 293, 407

\bibitem[]{} Georgelin Y.M. \& Georgelin Y.P, 1976, A\&A 49, 57

\bibitem[]{} Gies D. 2003, in proc, IAU Symp 212, {\it A Massive Star
Odyssey, from Main Sequence to Supernova}, 
eds. van der Hucht, K.A., Herrero A. \& Esteban C.. (San Francisco: ASP), p.~91

\bibitem[]{} Gr\"{a}fener G., Koesterke L. \& Hamann W-R., 2002, A\&A 387, 244

\bibitem[]{} Hanson M.M., Conti P.S. \& Rieke M.J. 1996, ApJS 107, 281

%\bibitem[\protect\citeauthoryear{{Haser}}{{Haser}}{1995}]{Haser:1995}
%{Haser}, S.~M. 1995.
%\newblock {\em Spektroscopie hei{\ss}er Sterne der Lokalen Gruppe im
%5  ultravioletten Spektralbereich}.
%\newblock Ph.\ D. thesis, Ludwig-Maximilians-Universit{\"a}t M{\"u}nchen

%\bibitem[]{} Harries T.J., Hilditch 1998, in Boulder-Munich II: Properties
%of Hot, Luminous Stars, ASP Conf. Ser 131, (ed. I.D. Howarth), p.401

\bibitem[]{} Haser S.M., Pauldrach A.W.A., Lennon D.J. et al.. 1998, 
A\&A 330, 285

\bibitem[Herrero et al. 1992]{1992A&A...261..209H} Herrero, A., Kudritzki, 
R. P., Vilchez, J. M., Kunze, D., Butler, K. \& Haser, S. 1992, A\&A, 261, 
209 

\bibitem[]{} Herrero A., Puls J. \& Najarro F., 2002, A\&A 396, 949

\bibitem[]{} Hilditch R.W., Harries T.J. \& Bell S.A., 1996, A\&A 314, 165

\bibitem[]{} Hillier D.J. \& Miller D.L. 1998, ApJ 496, 407

\bibitem[]{} Hillier D.J., Lanz T., Heap S.R. et al., 2003, ApJ 588, 1039

\bibitem[Howarth(1983)]{how83} Howarth I.D., 1983, MNRAS, 203, 301

\bibitem[]{} Howarth I.D., Siebert K.W., Hussain G.A.J. \& Prinja R.K.,
1997, MNRAS 284, 265

\bibitem[]{} Hubeny I. \& Lanz T., 2003, ApJS 146 in press (astro-ph/0210157)


\bibitem[]{} Humphreys R.M., 1978, ApJS 38, 309

\bibitem[]{} Kudritzki R.-P. \& Puls J., 2000, ARA\&A 38, 613

\bibitem[]{} Kudritzki R.-P., Pauldrach A.W.A. \& Puls J., 1987, A\&A 173, 293

\bibitem[]{KPPA} Kudritzki, R.-P., Pauldrach, A.W.A., Puls, J. \&
Abbott, D.C. 1989, A\&A 219, 205

\bibitem[]{} Kudritzki R.-P., Puls J., Lennon D.J. et al. 1999, A\&A 350, 970


\bibitem[]{} Lamers H.J.G.L.M. \& Leitherer C., 1993, ApJ 412, 771

\bibitem[]{} Lamers, H.J.G.L.M. \& Cassinelli J.P., 1999,  {\it Introduction to
Stellar Winds}, (Cambridge: CUP)

\bibitem[]{} Lamers H.J.G.L.M., Cerruti-Sola M. \& Perinotto M., 1987, ApJ 314, 726

\bibitem[]{} Lamers H.J.G.L.M., Snow T.P. \& Lindholm, D.M., 1995, ApJ 455, 269

\bibitem[]{} Lamers H.J.G.L.M., Harzevoort J.M.A.G., Schrijver H., Hoogerwerf
R. \& Kudritzki R.-P., 1997, A\&A 325, L25

\bibitem[]{} Leitherer C., Robert C. \& Drissen L., 1992, ApJ 401, 596

\bibitem[]{} Leitherer C., Chapman J.M. \& Koribalski B., 1995, ApJ 450, 289

\bibitem[]{} Lennon D.J., 1997, A\&A 317, 871
 
\bibitem[]{} Lucy L. \& White R.L., 1980, ApJ 241, 300

\bibitem[]{} Maeder A. \& Eenens P. (eds), 2003, Proc
IAU Symposium 215, {\it Stellar Rotation}, (San Francisco: ASP)

\bibitem[]{} Martins F., Schaerer D. \& Hillier D.J., 2002, A\&A 382, 999

\bibitem[]{} Massey P., Penny L.R. \& Vukovich J., 2002, ApJ 565, 982

\bibitem[]{} McErlean N.D., Lennon D.J. \& Dufton P.L., 1999, A\&A 349, 553

\bibitem[]{} Mihalas D., Hummer D.G. \& Conti P.S., 1972, ApJ 175, L99

\bibitem[]{} Morgan W.W., Keenan, P.C. \& Kellman E., 1943, {\it An Atlas of Stellar Spectra
with an Outline of Spectral Classification}, (Chicago: UCP)

\bibitem[]{} Morton, D.C., 1967, ApJ 150, 535

\bibitem[]{} Owocki S.P., Cranmer, S.R. \& Gayley K.G., 1996, ApJ 472, L115


\bibitem []{} Pauldrach A.W.A., Hoffman T.L. \& Lennon M., 2001, A\&A 375, 161

\bibitem[]{} Pellerin A., Fullerton A.W., Robert C. et al. 2002, ApJS 143, 159

\bibitem[]{} Petrenz  P. \& Puls J., 2000, A\&A 358, 956

\bibitem[]{} Pettini M., Steidel C., Adelberger K.L., Dickinson M. \& Giavalisco M., 2000, ApJ 528, 96

\bibitem[]{} Plaskett J.S. \& Pearce J.S., 1931, Publ. DAO 5, 1

\bibitem[]{} Prinja R.K., Barlow M.J. \& Howarth I.D., 1990, ApJ 361, 607

\bibitem[]{Prinja:1998} {Prinja}, R. K., \& {Crowther}, P. A. 1998, MNRAS,~300, 828

\bibitem[\protect\citeauthoryear{{Puls}, {Kudritzki}, {Herrero}, {Pauldrach},
  {Haser}, {Lennon}, {Gabler}, {Voels}, {Vilchez}, {Wachter}, and
  {Feldmeier}}{{Puls} et~al.}{1996}]{Puls:1996}
{Puls}, J., et~al. 1996, A\&A,~305, 171

%\bibitem[]{} Puls J. et al. 2002

\bibitem[]{} Runacres, M.C. \& Owocki, S.P., 2002, A\&A 381, 1015

\bibitem[]{} Santaloya-Rey A.E., Puls J. \& Herrero A., 1997, A\&A 323, 488

\bibitem[]{} Schaerer D. \& Schmutz  W., 1994, A\&A, 288, 231

\bibitem[Seaton(1979)]{mike79} Seaton M.J., 1979, MNRAS, 187, 73

\bibitem[]{} Smith L.F., Mezger, P.G. \& Biermann, P. 1978, A\&A, 66, 65

\bibitem[]{} Smith K.C. \& Howarth I.D., 1998, MNRAS 299, 1146

\bibitem[Taresch et al. 1997]{1997A&A...321..531T} Taresch, G., et al. 
1997, A\&A, 321, 531 

\bibitem[Vacca Garmany \& Shull 1996]{1996ApJ...460..914V} Vacca W.D., 
Garmany C.D. \& Shull J.M.  1996, ApJ, 460, 914 

\bibitem[]{} van der Hucht K.A., Herrero A. \& Esteban C. (eds), 2003, 
{\it IAU Symposium 212: A Massive Star Odyssey, from Main Sequence to Supernova, }
(San Francisco: ASP)


%Vink de Koter 2002 LBVs 393, 543

\bibitem[]{} Venn K., 1999, ApJ 518, 405

% bistability
\bibitem[]{} Vink J.S., de Koter A. \& Lamers H.J.G.L.M., 1999, A\&A 350, 181

% dM/dt(Z)
\bibitem[]{} Vink J.S., de Koter A. \& Lamers H.J.G.L.M., 2001, A\&A 369, 574

%\bibitem[]{} Walborn N.R., 1971, ApJ 167, L31 %(early O)

\bibitem[]{} von Zeipel H., 1924, MNRAS 84, 665


\bibitem[]{}Walborn, N.R., 1971a, ApJS 23, 257 % MK

\bibitem[]{}Walborn, N.R., 1971b, ApJ 167, L31 % early O

\bibitem[]{} Walborn N.R., 1976, ApJ 205, 419 % OBN, OBC

\bibitem[Walborn (1977)]{Walborn:1977}  {Walborn}, N.~R. 1977, ApJ~215, 53

\bibitem[]{} Walborn N.R., 2003, in proc, IAU Symp 212, {\it A Massive Star
Odyssey, from Main Sequence to Supernova},  eds. van der Hucht, K.A., Herrero A. \& Esteban C., 
(San Francisco: ASP), p.~13


% far-UV
%\bibitem[]{} Walborn N.R., Long K.S., Lennon D.J., Kudritzki R.-P., 1995, 
%ApJ 454 L27

\bibitem[]{} Walborn N.R., Lennon D.J., Haser S.M., Kudritzki R.-P. \& 
Voels S.A., 1995, PASP 107, 104

\bibitem[]{} Walborn N.R., Howarth I.D., Lennom D.J. et al., 2002a AJ  123, 2754 

\bibitem[]{} Walborn N.R., Fullerton A.W., Crowther P.A. et al. 2002b, ApJS 141, 443
%\bibitem[1994]{alref1} Aalto, S. \etal\  1994, A\&A, 286, 365.
%%% Using \cite{Bei} in the text
%\bibitem[1986]{Bei} Beichman, C.A., Neugebauer, G., Habing,
%   H., Clegg, P.E. \& Chester, T.C. 1988, editors, {\it ``IRAS Catalogs and
%   Atlases: Explanatory Supplement''}, NASA RP-1190 (Washington: NASA)
%\bibitem[1987]{ref1987} Beichman, C.A. 1987, ARA\&A, 25, 521
%\bibitem[1987]{so1987} Soifer, B.T., Houck, J.R. and Neugebauer, G. 1987, ARAA, 25, 187

\bibitem[]{} Wright A.E. \& Barlow M.J., 1975, MNRAS 170, 41


\end{thebibliography}
\end{document}